\documentclass[final,twocolumn]{pasj02} 

\jyear{2026}
\volume{xx}
\Issue{xx} 
\doi{psag093}
\Received{2026/04/18}
\Accepted{2026/07/03}
\Published{2026/08/06}

\begin{document} 

\title{Radio Constraints on the Circumstellar Environment of the Type IIb Supernova SN\,2024iss}

\author{
 Yuhei \textsc{Iwata},\altaffilmark{1, 2}\altemailmark\orcid{0000-0002-9255-4742} \email{yuhei.iwata@nao.ac.jp}
Tomoki \textsc{Matsuoka},\altaffilmark{3, 4, 5}\orcid{0000-0002-6916-3559}
Masanori  \textsc{Akimoto},\altaffilmark{6, 7}
Keiichi \textsc{Maeda},\altaffilmark{8}\orcid{0000-0003-2611-7269}
Nozomu \textsc{Tominaga},\altaffilmark{2,9,10}\orcid{0000-0001-8537-3153}
Yoshinori \textsc{Yonekura},\altaffilmark{11}\orcid{0000-0001-5615-5464}
Takashi J. \textsc{Moriya},\altaffilmark{2,9,12}\orcid{0000-0003-1169-1954}
Kotaro  \textsc{Niinuma}\altaffilmark{6,7}\orcid{0000-0002-8169-3579}
and 
Kenta \textsc{Fujisawa},\altaffilmark{6, 7}\orcid{0009-0008-1070-4411}}

\altaffiltext{1}{Mizusawa VLBI Observatory, National Astronomical Observatory of Japan, 2-12 Hoshigaoka, Mizusawa, Oshu, Iwate 023-0861, Japan}
\altaffiltext{2}{Astronomical Science Program, Graduate Institute for Advanced Studies, SOKENDAI, 2-21-1 Osawa, Mitaka, Tokyo, 181-8588, Japan}
\altaffiltext{3}{Department of Earth Science and Astronomy, Graduate School of Arts and Sciences, The University of Tokyo, Tokyo 153-8902, Japan}
\altaffiltext{4}{Department of Physics, National Chung Hsing University, No. 145, Xingda Rd., South Dist., Taichung 40227, Taiwan}
\altaffiltext{5}{Institute of Astronomy and Astrophysics, Academia Sinica, No.1, Sec.4, Roosevelt Road, Taipei 106319, Taiwan}
\altaffiltext{6}{The Research Institute for Time Studies, Yamaguchi University, 1677-1 Yoshida, Yamaguchi-city, Yamaguchi 753-8511, Japan}
\altaffiltext{7}{Graduate School of Science and Technology for Innovation, Yamaguchi University, 1677-1 Yoshida, Yamaguchi-city, Yamaguchi 753-8512, Japan}
\altaffiltext{8}{Department of Astronomy, Kyoto University, Kitashirakawa-Oiwake-cho, Sakyo-ku, Kyoto 606-8502, Japan}
\altaffiltext{9}{Division of Science, National Astronomical Observatory of Japan, 2-21-1 Osawa, Mitaka, Tokyo 181-8588, Japan}
\altaffiltext{10}{Department of Physics, Faculty of Science and Engineering, Konan University, 8-9-1 Okamoto, Kobe, Hyogo, 658-8501, Japan}
\altaffiltext{11}{Center for Astronomy, Ibaraki University, 2-1-1 Bunkyo, Mito, Ibaraki 310-8512, Japan}
\altaffiltext{12}{School of Physics and Astronomy, Faculty of Science, Monash University, Clayton, Victoria 3800, Australia}



\maketitle

\begin{abstract}
Type~IIb supernovae exhibit diverse progenitor properties, and radio observations offer a unique probe of their mass-loss histories shortly before the explosion. We present Japanese VLBI Network single-baseline monitoring of the nearby Type~IIb SN\,2024iss at 6.9 and 8.4\>GHz, spanning approximately one year after its discovery. Our radio observations have detected its emission at 10 and 23 days after the explosion, with subsequent epochs yielding non-detections. Based on the peak radio luminosity and peak time, SN\,2024iss exhibits radio properties highly comparable to those of compact-envelope events. Using a synchrotron self-absorption (SSA) modeling, we estimate a progenitor mass-loss rate of $\dot{M} \approx 2.5 \times 10^{-6}\>M_{\odot}\>{\rm yr^{-1}}$ for a compact progenitor wind velocity of $100 \>{\rm km\>s^{-1}}$. Furthermore, our SSA analysis yields a mean expansion velocity of $V_{\rm sh} \approx 3.3 \times 10^4\>{\rm km\>s^{-1}}$, which exceeds the theoretical shock velocity derived from the self-similar solution by a factor of $\sim 2.4$. Even for the conservative upper-bound peak time, the SSA-derived velocity remains larger than the theoretical expectation by a factor of $\gtrsim 1.7$. To explain this velocity excess, we propose the presence of a confined, dense circumstellar matter (CSM) surrounding the progenitor. The shock emergence from this confined CSM may have accelerated the forward shock, pointing to a highly complex and non-steady mass-loss history shortly before the explosion.
\end{abstract}

\begin{center}
\begin{minipage}{0.95\columnwidth}
\footnotesize
This is a pre-copyedited, author-produced version of an article accepted for publication in ``Publications of the Astronomical Society of Japan'' following peer review. The version of record is available online at [https://doi.org/10.1093/pasj/psag093].
\end{minipage}
\end{center}


\section{Introduction}

Type~IIb supernovae (SNe~IIb) are core-collapse explosions characterized observationally by hydrogen features at early phases that fade with time, with the spectra evolving to resemble those of hydrogen-poor SNe~Ib/Ibc (e.g., \cite{Filippenko93, Nomoto93}). This spectroscopic transition indicates that only a small amount of hydrogen remains at the time of explosion, implying that the progenitor has been stripped of most of its hydrogen-rich envelope prior to core collapse. It is widely accepted that such partial stripping is predominantly driven by mass transfer to a binary companion (e.g., \cite{Podsiadlowski92, Nomoto95, Yoon10, Claeys11, Benvenuto13}). Alternatively, the envelope could also be lost via radiatively driven stellar winds (e.g., \cite{Smith08}). Therefore, SNe~IIb provide important clues to the late-stage evolution of massive stars and mass transfer in interacting binaries.

A growing number of well-observed SNe~IIb have enabled direct and indirect constraints on their progenitors, revealing significant diversity in their pre-explosion states. In some cases, observations have provided progenitor identifications in pre-explosion images (e.g., \cite{Aldering94, VanDyk11, VanDyk14}), as well as a companion star in the binary system \citep{Folatelli14}. Furthermore, early-time infrared, optical and ultraviolet observations, along with bolometric light-curve modeling, have offered complementary estimates of the progenitor radius and the residual-envelope structure (e.g., \cite{Bersten12, Bufano14}). These multi-wavelength constraints suggest that SNe~IIb arise from progenitors with a wide range of stellar radii and envelope configurations, pointing to variations in the stripping processes and the final mass-loss history.

The mass loss prior to explosion shapes the circumstellar matter (CSM) surrounding the progenitor. As the rapidly expanding SN shock interacts with this CSM, relativistic electrons and amplified magnetic fields produce non-thermal synchrotron radiation (e.g., \cite{Chevalier98, Weiler02}). Because the radio light curve is initially shaped by absorption processes such as free-free absorption and synchrotron self-absorption, its temporal evolution encodes crucial information about the shock velocity and the CSM density. Consequently, radio monitoring provides a direct and unique probe of the progenitor's mass-loss rate on timescales from years up to centuries prior to the explosion (e.g., \cite{Iwata25}).

Extensive radio monitoring of SNe~IIb has revealed their distinct light-curve properties. \citet{Bietenholz21} reported that the mean peak spectral luminosity of SNe~IIb ($10^{26.3}$--$10^{27.3}\>{\rm erg\>s^{-1}\>Hz^{-1}}$) is relatively high and has a narrower distribution compared to that of all SNe~II ($10^{24.0}$--$10^{26.6}\>{\rm erg\>s^{-1}\>Hz^{-1}}$). By modeling these radio light curves, the mass-loss rates of their progenitors have been derived, revealing a wide range of values. For instance, the well-studied SN\,1993J was suggested to have a relatively high mass-loss rate of $(2-6) \times 10^{-5}\>M_{\odot}\>\mathrm{yr}^{-1}$ \citep{Fransson96}, whereas SN\,2008ax exhibited a lower mass-loss rate of $(1-6) \times 10^{-6}\>M_{\odot}\>\mathrm{yr}^{-1}$ \citep{Roming09}, with both assuming a wind velocity of $10\>{\rm km\>s^{-1}}$. These observations underscore the diversity in the mass-loss histories among SNe~IIb.

SN\,2024iss was discovered on 2024 May 12 21:37 UT as a nearby bright transient \citep{ONeill24} and was subsequently classified as a Type~IIb SN \citep{Srivastav24}. Based on the location, the nearby dwarf galaxy WISEA J125906.48+284842.6 (redshift $z = 0.003334$, corresponding distance $D \approx 14.1$\>Mpc) is considered its host galaxy. Early-time ultraviolet, optical, and near-infrared observations and light-curve modeling suggest that its progenitor could be a blue or yellow supergiant \citep{Yamanaka25}. Because of its proximity, SN\,2024iss is an excellent target to trace its radio characteristics, and thereby to understand the final evolution of the progenitor star.

This paper is organized as follows. In section~\ref{sec:2}, we present the Japanese VLBI Network (JVN) single-baseline observations of SN\,2024iss at 6.9 and 8.4\>GHz, spanning approximately one year after discovery, along with the data analysis procedures. In section~\ref{sec:3}, we report the obtained radio flux densities and the temporal evolution of the spectral luminosity. In section~\ref{sec:4}, we derive the progenitor mass-loss rate based on the synchrotron emission model, compare the radio properties of SN\,2024iss with those of other SNe~IIb such as SN\,2008ax, and discuss the nature of its progenitor. Finally, we summarize our conclusions in section~\ref{sec:5}.

\section{Observations and Data Analysis}\label{sec:2}
The JVN observations were conducted as a single baseline VLBI using two radio telescopes: Hitachi 32\>m (Hit32, \cite{Yonekura16}) and Yamaguchi 32\>m (Yam32, \cite{Fujisawa22}). The observations were conducted in 9 epochs spanning from 3 days to 366 days after the discovery report on SN\,2024iss. The observation intervals were set to be roughly equal on the logarithmic scale, except for the Yam32 maintenance period from 2024 August to September. The observation settings are the same as those of SN\,2023ixf \citep{Iwata25}. Simultaneous observations were made on two frequency bands, one centered on 6.856\>GHz (C band) and the other on 8.448\>GHz (X band), each with a bandwidth of 512\>MHz. Due to an issue with the local oscillator of Yam32, observations in the last three epochs were carried out with only one frequency band. Left-hand circular polarizations were received and sampled with 2-bit quantization. The distance between Hit32 and Yam32 is 873\>km, corresponding to angular resolutions of 10\>mas at the C band and 8.4\>mas at the X band. We observed the quasar 3C286 as a fringe finder and the ICRF3 source J130028.5+283010 (J1300+2830) as a gain calibrator. The on-source integration times for 3C286, J1300+2830, and SN\,2024iss were 10\>min, 5\>min, and 10\>min, respectively. The target and gain calibrator observations were repeated two times.

Data correlation was performed using the FX-type software correlator GICO3, developed by the National Institute of Information and Communications Technology. We first measured a clock delay and rate by a fringe search for 3C286. Adopting the delay and rate of 3C286, we performed fringe searches for J1300+2830 iterating three times to obtain the accurate delay and rate. We searched for any fringe peak for SN\,2024iss close to the delay and rate estimated by the gain calibrator (delay within $\pm\, 10$\>ns and rate within $\pm\, 15$\>mHz). We regarded a detection when the fringe peaks from the two scans and two frequency bands (i.e., four peaks) were consistently located at the same position in the delay--rate diagram. Since the coherence times of 6.9 and 8.4\>GHz are slightly shorter than the integration time of 10\>min, we searched for the integration times ($T_{\rm integ}$) that make the signal-to-noise ratios of the emission highest, and used them for the subsequent analysis. When we could not detect any fringe signal, we set a 5\,$\sigma$ upper limit to the first scan.

Flux density scaling was performed by using the flux density of J1300+2830 measured by the Yamaguchi Interferometer (YI, \cite{Fujisawa22}), which is a connected array consisting of Yamaguchi 32\>m and 34\>m radio telescopes located 108\>m apart. The flux scaling of YI was performed by the stable flux calibrator 3C286 and assumed its flux density from \citet{Perley17} with 5\% uncertainty. Observations of YI were carried out on 2024 May 21 and 2025 May 14, and the measured flux densities of J1300+2830 at 6.9 and 8.4\>GHz were $272.7\,\pm\,13.8$\>mJy and $257.2\,\pm\,14.2$\>mJy on 2024 May 21, and $255.2\,\pm\,12.8$\>mJy and $243.5\,\pm\,13.4$\>mJy on 2025 May 14. We adopted the former flux densities to the first 5 epochs of the Hit32-Yam32 observations, and the latter to the last 4 epochs. Since the observations of SN\,2024iss and the gain calibrator of J1300+2830 were conducted at almost the same elevation angle, we ignored the dependence of aperture efficiency on the elevation angle. The uncertainties of the flux densities for individual scans were derived by propagating the 1\,$\sigma$ statistical errors of the correlated amplitudes and the uncertainties of the flux scaling. For the epochs with detections, the flux densities of the two scans were averaged, and their final uncertainties were estimated by combining the propagated errors and the scatter between the scans.

\begin{table*}
  \tbl{JVN observations.}{%
  \begin{tabular}{cccccccc}
      \hline
      Obs. code & Date and Time\footnotemark[$*$] & MJD & $t_{\rm exp}$\footnotemark[$\dagger$] & $T_{\rm integ, 6.9 GHz}$ & $F_{\rm 6.9 GHz}$\footnotemark[$\ddagger$] & $T_{\rm integ, 8.4 GHz}$  & $F_{\rm 8.4 GHz}$\footnotemark[$\ddagger$]\\ 
       & (UT) &  & (days) & (seconds) & (mJy) & (seconds)  & (mJy)\\ 
      \hline
U24136A & 2024-05-15 09:21:12 & 60445.39 & 2.99 & 600 & $<$3.1 & 600 & $<$3.0 \\\hline
U24143A & 2024-05-22 08:53:40 & 60452.37 & 9.97 & 440 & $3.0\,\pm\,0.7$ & 380 & $3.3\,\pm\,0.8$  \\
  & 2024-05-22 09:10:40 & 60452.38 & 9.98 & 580 & $2.4\,\pm\,0.6$ & 410 & $3.1\,\pm\,0.7$  \\
average & 2024-05-22 09:04:40 & 60452.38 & 9.98 &   & $2.7\,\pm\,0.6$ &   & $3.2\,\pm\,0.5$  \\\hline
U24156A & 2024-06-04 08:02:32 & 60465.34 & 22.93 & 460 & $3.6\,\pm\,0.8$ & 380 & $2.7\,\pm\,0.8$  \\
  & 2024-06-04 08:19:32 & 60465.35 & 22.94 & 410 & $3.8\,\pm\,0.8$ & 400 & $3.9\,\pm\,0.8$  \\
average & 2024-06-04 08:13:32 & 60465.34 & 22.94 &   & $3.7\,\pm\,0.6$ &   & $3.3\,\pm\,0.8$  \\\hline
U24179A & 2024-06-27 06:32:04 & 60488.27 & 45.87 & 600 & $<$3.4 & 600 & $<$3.3 \\
U24291A & 2024-10-16 23:11:32 & 60599.97 & 157.56 & 600 & $<$4.4 & 600 & $<$3.6 \\
U24340A & 2024-12-05 19:54:52 & 60649.83 & 207.43 & 600 & $<$2.9 & 600 & $<$2.2 \\
U25034A & 2025-02-03 15:58:52 & 60709.67 & 267.26 &   &   & 600 & $<$2.5 \\
U25049A & 2025-02-18 14:59:52 & 60724.62 & 282.22 & 600 & $<$2.9 &  &   \\
U25133A & 2025-05-13 09:29:28 & 60808.40 & 365.99 & 600 & $<$3.0 &  &   \\\hline
    \end{tabular}}\label{tab:1}
\begin{tabnote}
\footnotemark[$*$] Dates and times are at the midpoints of the observations. \\
\footnotemark[$\dagger$] Time since explosion ($t_{\rm exp}$) is given relative to an estimated explosion time of MJD = 60442.40 \citep{Yamanaka25}. \\
\footnotemark[$\ddagger$] Non-detections: flux-density upper limits correspond to $5\sigma$ of the correlated amplitude. Detections: flux-density uncertainties for individual scans at each detected epoch correspond to the $1\sigma$ correlated-amplitude uncertainties combined with the flux-scaling errors; uncertainties in the averaged flux densities are obtained by combining the propagated errors with the scatter between the two scans (see section~\ref{sec:2}).
\end{tabnote}
\end{table*}

\section{Results}\label{sec:3}

\begin{figure}
 \begin{center}
  \includegraphics[width=8cm]{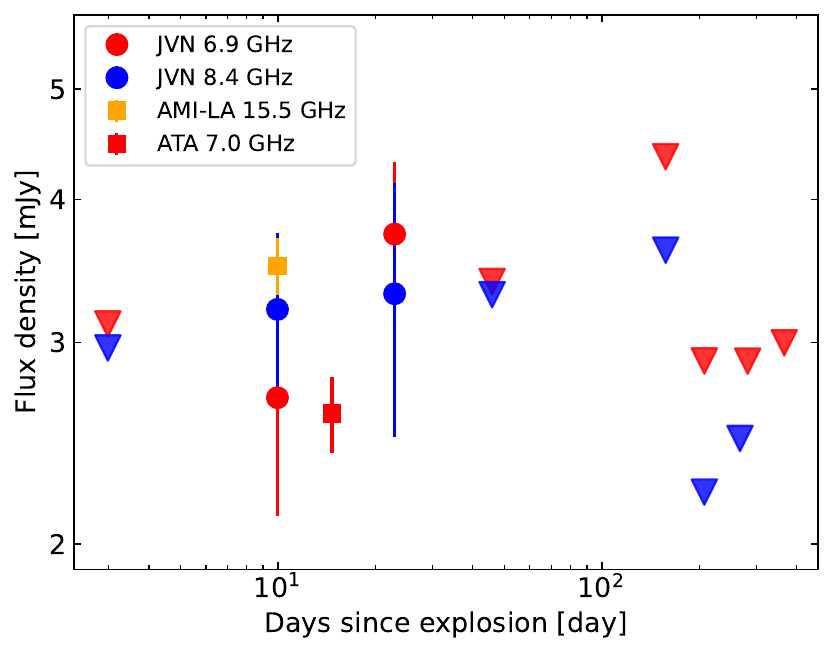} 
 \end{center}
\caption{Observed radio flux densities of SN\,2024iss. Circles and triangles indicate JVN-detected flux densities and 5\,$\sigma$ upper limits, respectively. Squares are the reported flux densities observed by AMI-LA at 15.5\>GHz \citep{Sfaradi24} and by ATA at 7.0\>GHz \citep{Bright24}.
 {Alt text: A graph showing the time evolution of flux density after a supernova explosion. The x-axis represents days since the explosion from 2 to 500 days, and the y-axis shows observed flux density from 2 to 5 millijansky.}
}\label{fig:1}
\end{figure}

The measured flux densities of SN\,2024iss by JVN are summarized in table~\ref{tab:1}, and shown in figure~\ref{fig:1}. The time since explosion ($t_{\rm exp}$) of each observation was calculated using the estimated SN first light of MJD = 60442.40 \citep{Yamanaka25} from the first detection on MJD = 60442.90 and the last non-detection on MJD = 60441.90 \citep{ONeill24}. Two radio flux densities were reported for SN\,2024iss, $3.5\,\pm\,0.2$\>mJy at 15.5\>GHz on 2024 May 22 ($t_{\rm exp} \sim 10$\>days) by the Arcminute Microkelvin Imager - Large Array (AMI-LA, \cite{Sfaradi24}) and $2.6\,\pm\,0.2$\>mJy at 7.0\>GHz on 2024 May 27 ($t_{\rm exp} = 14.7$\>days) by the Allen Telescope Array (ATA, \cite{Bright24}). Our JVN detections on 2024 May 22 and 2024 June 4 are consistent with them.

\begin{figure}
 \begin{center}
  \includegraphics[width=8cm]{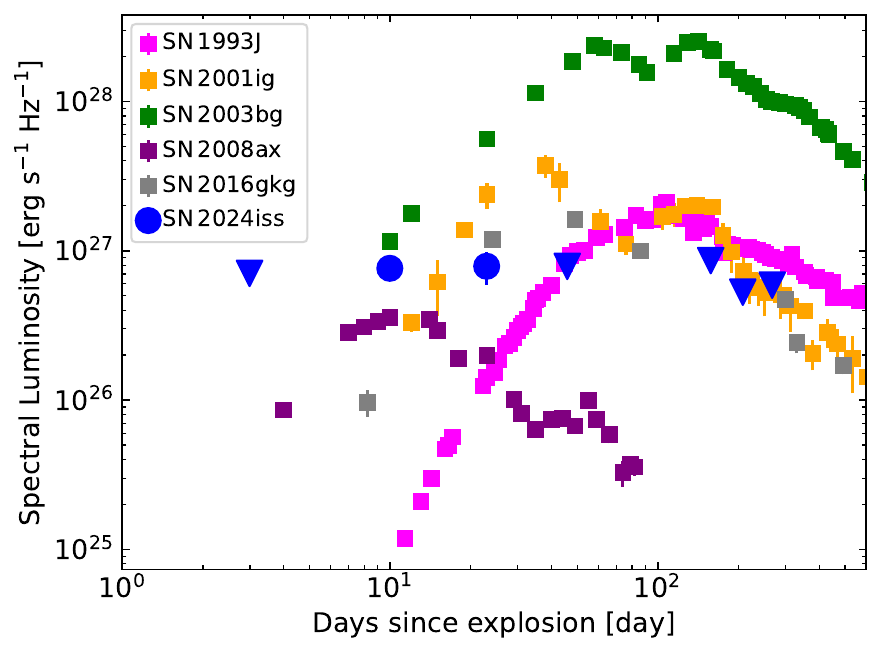} 
 \end{center}
\caption{Spectral luminosity evolution of SN\,2024iss compared with other well-observed SNe~IIb (magenta: SN\,1993J from \cite{Weiler07}, orange: SN\,2001ig from \cite{Ryder04}, green: SN\,2003bg from \cite{Soderberg06}, purple: SN\,2008ax from \cite{Roming09} with an assumed distance of 7.77 Mpc provided by the NASA/IPAC Extragalactic Database,  grey: SN\,2016gkg from \cite{Nayana22}). All measurements were in the 8\>GHz band.
{Alt text: A graph comparing the spectral luminosity evolution of six Type~IIb supernovae. The x-axis shows days since explosion from 1 to 600 days, and the y-axis indicates spectral luminosity in erg per second per hertz. Each color represents a different supernova: SN 1993J, SN 2001ig, SN 2003bg, SN 2008ax, SN 2016gkg, and SN 2024iss for this measurement.}}\label{fig:2}
\end{figure}

Figure~\ref{fig:2} shows the luminosity evolution of SN\,2024iss in the 8.4\>GHz band in comparison to those of well-observed other SNe~IIb. The detected spectral luminosity of $\sim 10^{27} \,{\rm erg\>s^{-1}\>Hz^{-1}}$ is comparable to the mean peak spectral luminosity of $L_{\rm p} = 10^{26.8 \,\pm\, 0.5}\,{\rm erg\>s^{-1}\>Hz^{-1}}$ for SNe~IIb \citep{Bietenholz21}. The luminosity peak of SN\,2024iss is around two detections at 10 and 23 days. This timescale to reach the peak ($t_{\rm p}$) is relatively short compared to the typical value of $10^{1.5\,\pm\,0.6}$\>days for SNe~IIb \citep{Bietenholz21} but still within the range of the variation.

The light curve of SN\,2008ax has similar $L_{\rm p}$ and $t_{\rm p}$ as that of SN\,2024iss. The estimated mass-loss rate of SN\,2008ax is $(1-6) \times 10^{-6}\>M_{\odot}\>\mathrm{yr}^{-1}$ \citep{Roming09}, and the mass-loss rate of SN\,2024iss would be similar to it. Some SNe~IIb show multiple peaks in their light curves, but SN\,2024iss did not show another peak with comparable or higher luminosity during our observational period. More sensitive radio telescopes potentially can detect another peak similar to SN\,2008ax.

\section{Discussion}\label{sec:4}

\subsection{Synchrotron Self-absorption Modeling}
To investigate the properties of the circumstellar environment and the progenitor of SN\,2024iss, we first determine its location in the peak radio spectral luminosity ($L_{\rm p}$) versus peak time ($t_{\rm p}$) diagram. According to \citet{Chevalier10}, SNe~IIb can be divided into two populations: extended-envelope (eIIb) and compact-envelope (cIIb) events. In the $L_{\rm p}$--$t_{\rm p}$ diagram, eIIb and cIIb events can generally be distinguished by their different locations. Given the limited sampling around the radio maximum, we do not fit for the peak parameters and instead adopt the 6.9\>GHz flux density at 22.94\>days ($F_{\rm 6.9\>GHz}=3.7\,\pm\,0.6$\>mJy) as a representative peak value, yielding $L_{\rm p} \approx 8.8 \times 10^{26}\>{\rm erg\>s^{-1}\>Hz^{-1}}$ and the 5\>GHz scaled peak time of $t_{\rm p, 5\>GHz} \approx 31.5$\>days. The physical quantities derived below from this representative peak are therefore treated as approximate estimates.

\begin{figure*}
 \begin{center}
  \includegraphics[width=14cm]{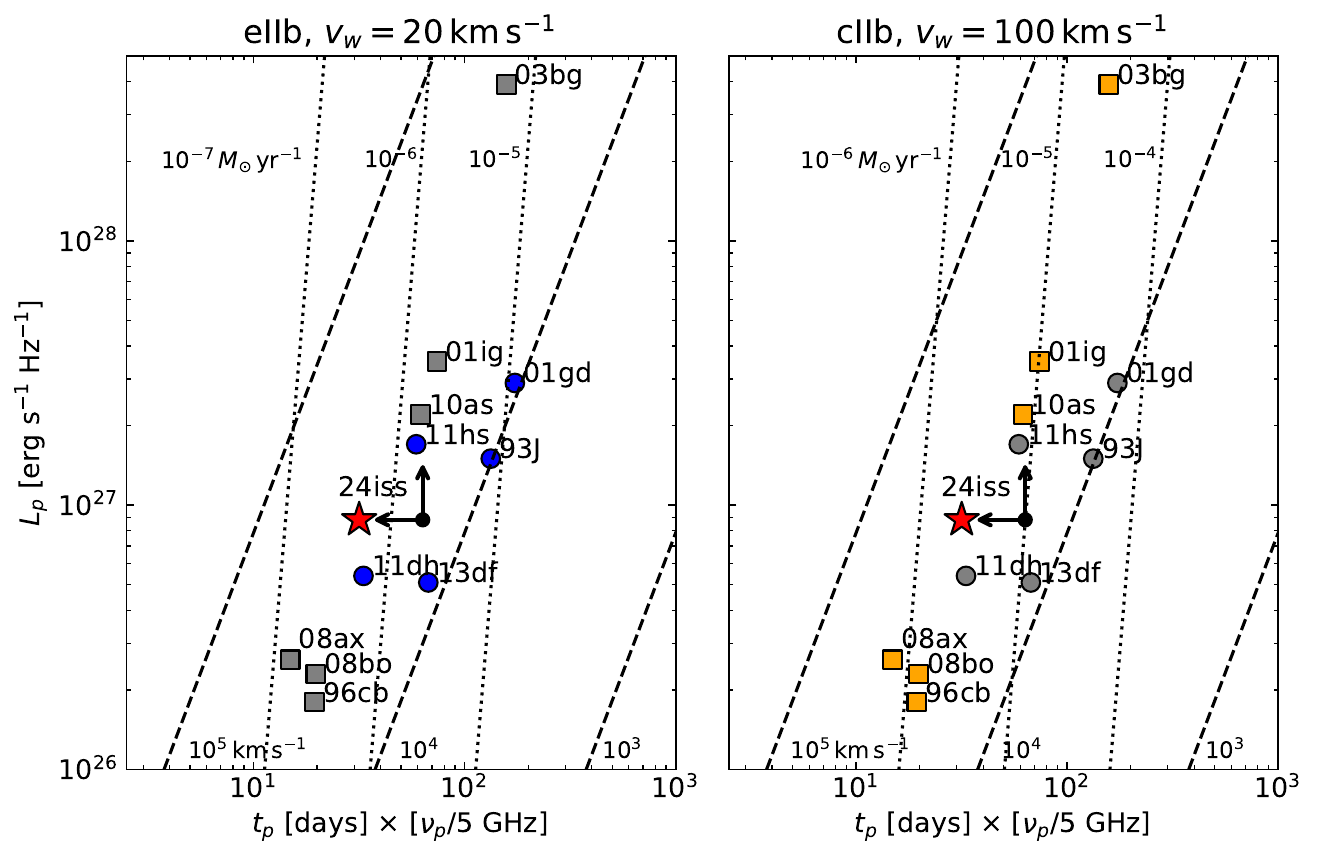} 
 \end{center}
\caption{Peak radio spectral luminosities vs.\ time of peak of the light curves for SNe~IIb. The properties of SN\,2024iss are compared with eIIb (blue circles) in the left panel, and cIIb (orange squares) in the right panel. The observed SNe are designated by the last two digits of the year and letters. The dashed lines indicate the mean velocity of the radio shell if SSA is responsible for the flux peak; a particle spectral index $p=3$ is assumed (see \cite{Chevalier98, Chevalier06}). The dotted lines represent the mass-loss rates calculated using the equations of \citet{Chevalier06} with assumed wind velocities of $20\>{\rm km\>s^{-1}}$ for eIIb and $100\>{\rm km\>s^{-1}}$ for cIIb, and $\epsilon_e = \epsilon_B = 0.1$. The arrows indicate the upper limit of the peak time and lower limit of the peak luminosity of SN\,2024iss.
 {Alt text: Scatter plot of peak radio spectral luminosity versus the time of peak for Type~IIb supernovae. The x-axis shows the product of the time of the peak and the frequency of the measurement, and the y-axis shows peak spectral luminosity (erg\>s$^{-1}$\>Hz$^{-1}$). Dashed diagonal lines indicating the mean velocities of the radio shell and dotted diagonal lines representing the mass-loss rates are overplotted.}}
\label{fig:l-t}
\end{figure*}

Figure~\ref{fig:l-t} presents the $L_{\rm p}$--$t_{\rm p}$ diagram, in which we add SN\,2024iss to a subset of the compilation by \citet{Nayana22} that includes only objects with robust cIIb or eIIb classifications. For reference, pre-explosion imaging studies estimated a progenitor radius of $\sim 600\>R_{\odot}$ for SN\,1993J \citep{Aldering94}, which forms the basis for its classification as an eIIb event, whereas SN\,2008ax had a much smaller estimated radius of $30$--$50\>R_{\odot}$ \citep{Folatelli15}, leading to its classification as a cIIb event. In this diagram, SN\,2024iss lies on the left side, a region typically populated by cIIb events such as SN\,2008ax.

Based on optical, ultraviolet, and X-ray observations, \citet{Chen26} suggested that SN\,2024iss had a progenitor radius of $R = 244\,\pm\,43\>R_{\odot}$, which is larger than that of SN\,2008ax but smaller than that of SN\,1993J. Following the framework of \citet{Maeda15}, they interpreted SN\,2024iss as a transitional event between the cIIb and eIIb populations. This interpretation seems inconsistent with our radio-based classification, which indicates a cIIb event resembling SN\,2008ax. However, similar discrepancies between radio properties and multi-wavelength or pre-explosion inferences have been reported in other SNe~IIb. For SN\,2011dh, the radio emission and early bolometric light curve suggested a compact progenitor \citep{Arcavi11, Soderberg12}, whereas pre-explosion imaging identified an extended yellow supergiant (YSG) with a radius of $\sim 200\>R_{\odot}$ \citep{Maund11, VanDyk11, Bersten12}. Although this YSG was initially debated to be a binary companion, its post-explosion disappearance confirmed it as the progenitor \citep{VanDyk13}. Similarly, for SN\,2011hs, the progenitor radius inferred from the bolometric light curve suggests an eIIb progenitor, while its radio properties favor a cIIb classification \citep{Bufano14}.

Assuming the radio peak is governed by synchrotron self-absorption (SSA), we apply the formulation of \citet{Chevalier98, Chevalier06} to derive the forward shock radius and mean forward shock velocity at the time of the SSA peak (see dashed lines in figure~\ref{fig:l-t}). Here we set the microphysics parameters to $\epsilon_e = 0.1$ and $\epsilon_B = 0.1$. From the representative peak values, we derive the radius of the emission region to be $R_{\rm p} \approx 9.1 \times 10^{15}$\>cm, which corresponds to the mean shock velocity of $V_{\rm sh,\ SSA} \approx R_{\rm p}/t_{\rm p, 5\>GHz} \approx 3.3 \times 10^4 \>{\rm km\>s^{-1}}$.

Based on this SSA modeling \citep{Chevalier06}, we can also estimate the progenitor's mass-loss rate from the peak luminosity and peak time (indicated by the dotted lines in figure~\ref{fig:l-t}). Assuming a wind velocity of $v_{\rm w} = 100\>{\rm km\>s^{-1}}$ for a cIIb progenitor and $v_{\rm w} = 20\>{\rm km\>s^{-1}}$ for an eIIb progenitor, the mass-loss rates are derived to be $\dot{M} \approx 2.5 \times 10^{-6}\>M_{\odot}\>{\rm yr^{-1}}$ and $\dot{M} \approx 5.0 \times 10^{-7}\>M_{\odot}\>{\rm yr^{-1}}$, respectively. This latter value for the eIIb scenario is substantially lower than those of typical SNe~IIb, although we note that this estimate strongly depends on the assumed wind velocity and microphysics parameters.

\subsection{Shock Velocity Comparison}
In the previous section, we assumed that the radio emission peak is described solely by SSA. To validate this assumption, we check the effect of free-free absorption (FFA). The optical depth for FFA ($\tau_{\rm FFA}$) can be expressed as
\begin{eqnarray}
\tau_{\rm FFA} =&& 0.98 \times \left(\frac{T_e}{10^4\>{\rm K}}\right)^{-3/2} \left(\frac{\dot{M}}{10^{-5}\>M_{\odot}\>{\rm yr^{-1}}}\right)^{19/8} \nonumber \\
&&\times \left(\frac{v_{\rm w}}{10^7\>{\rm cm\>s^{-1}}}\right)^{-19/8} \left(\frac{E_{\rm kin}}{0.94 \times 10^{51}\>{\rm erg}}\right)^{-21/16} \nonumber \\
&&\times \left(\frac{M_{\rm ej}}{2.8\>M_{\odot}}\right)^{15/16} \left(\frac{t}{10\>{\rm days}}\right)^{-21/8} \left(\frac{\nu}{10\>{\rm GHz}}\right)^{-2},
\end{eqnarray}
where $T_e$ is the electron temperature of the CSM, $E_{\rm kin}$ is the kinetic energy of the ejecta, and $M_{\rm ej}$ is the ejecta mass. Assuming $T_e = 10^4$\>K, and adopting $E_{\rm kin} = 0.94 \times 10^{51}$\>erg and $M_{\rm ej} = 2.8\>M_{\odot}$ inferred for SN\,2024iss \citep{Yamanaka25}, we evaluate the optical depth using the derived mass-loss rate for the cIIb case ($\dot{M} \approx 2.5 \times 10^{-6} \> M_{\odot} \> {\rm yr^{-1}}$ and $v_{\rm w} = 100 \> {\rm km\>s^{-1}}$). At the representative peak epoch ($t_{\rm exp} = 22.94$\>days) and observing frequency ($\nu = 6.9$\>GHz), we calculate $\tau_{\rm FFA} \approx 9 \times 10^{-3}$. Since $\tau_{\rm FFA} \ll 1$, FFA is negligible at the peak epoch, confirming that the peak is indeed governed by SSA.

To further investigate the shock dynamics, we compare the mean expansion velocity derived from the SSA analysis ($V_{\rm sh,\ SSA} \approx 3.3 \times 10^4\>{\rm km\>s^{-1}}$) with the theoretical shock velocity expected from the self-similar solution \citep{Chevalier82a, Chevalier82b}. The expected shock velocity ($V_{\rm sh,\ theory}$) for a cIIb-like compact progenitor, assuming an outer ejecta density profile index of $n=10$, is given by
\begin{eqnarray}
V_{\rm sh,\ theory} =&& 1.32 \times 10^4\>{\rm km\>s^{-1}} \left(\frac{\dot{M}}{10^{-5}\>M_{\odot}\>{\rm yr^{-1}}}\right)^{-1/8} \nonumber \\
&&\times \left(\frac{v_{\rm w}}{100\>{\rm km\>s^{-1}}}\right)^{1/8} \left(\frac{E_{\rm kin}}{0.94 \times 10^{51}\>{\rm erg}}\right)^{7/16} \nonumber \\
&&\times \left(\frac{M_{\rm ej}}{2.8\>M_{\odot}}\right)^{-5/16} \left(\frac{t}{10\>{\rm days}}\right)^{-1/8}.
\label{eq:v_sh}
\end{eqnarray}
To be consistent with the shock velocity evaluated by the SSA modeling from the 5\>GHz peak, we calculate this theoretical velocity at the scaled peak time of $t_{\rm p, 5\>GHz} \approx 31.5$ days, yielding $V_{\rm sh,\ theory} \approx 1.4 \times 10^4\>{\rm km\>s^{-1}}$. The velocity derived from the SSA modeling is thus larger by a factor of $\sim 2.4$ compared to this theoretical expectation. We note that changing the outer ejecta density profile index does not substantially change this estimation.

Since the above discussion is based on the assumption that the detection at 22.9\>days corresponds to the 6.9\>GHz peak, we also evaluate the conservative case in which the peak time is bounded by the non-detection at 45.9\>days. Adopting this upper-bound timescale, $t_{\rm upp, 5\>GHz} = 62.9$\>days, together with $L_{\rm p} \approx 8.8 \times 10^{26}\>{\rm erg\>s^{-1}\>Hz^{-1}}$, which can be considered as a lower limit on the peak luminosity, gives a conservative lower limit of $V_{\rm sh,\ SSA} > 1.7 \times 10^4 \>{\rm km\>s^{-1}}$ and a cIIb mass-loss-rate upper limit of $\dot{M}<1.0\times10^{-5}\>M_{\odot}\>{\rm yr^{-1}}$ (see black circle in figure~\ref{fig:l-t}). Using equation (\ref{eq:v_sh}), this mass-loss-rate upper limit gives $V_{\rm sh,\ theory} \simeq 1.0 \times 10^4\>{\rm km\>s^{-1}}$, so that the velocity ratio remains $\gtrsim 1.7$. We note that this ratio, calculated by the upper limit of the peak time and the lower limit of the peak luminosity, is the most conservative value because
\begin{equation}
\frac{V_{\rm sh,\ SSA}}{V_{\rm sh,\ theory}} \propto L_{\rm p}^{17/38}t_{\rm p}^{-5/8}. 
\end{equation}

To explain the high shock velocity derived from the SSA modeling, we discuss the possible presence of a confined, dense CSM around the progenitor. Theoretical studies have demonstrated that when a forward shock emerges from a confined CSM into a lower-density environment, the shock velocity can increase by a factor of roughly two compared to that calculated from the self-similar solution for a steady wind \citep{Matsuoka19, Matsuoka25}. Furthermore, early X-ray spectral observations of SN\,2024iss \citep{Chen26} independently suggest the existence of such a confined CSM within a radius of $1.3 \times 10^{14}$\>cm. This scale is consistent with the larger emission radius derived from our SSA modeling ($R_{\rm p} \approx 9.1 \times 10^{15}$\>cm), suggesting that the radio observations may trace the accelerated shock after passing through this dense region. These consistent multi-wavelength findings support a scenario of a non-steady mass-loss history for the progenitor of SN\,2024iss shortly before the explosion.

Observational studies proposing a confined, dense CSM specifically within the Type~IIb population are still limited. Within this limited sample, early X-ray spectroscopy of SN\,2011dh suggested a circumstellar density enhancement \citep{Sasaki12}, although a later radio and X-ray long-term monitoring concluded that its CSM is more naturally explained by a nearly steady wind \citep{Maeda14}. Also, the possibility of a confined CSM is discussed for SN\,2018ivc, which has been proposed as a transitional event between Types~IIL and IIb \citep{Maeda23}. Adding to this limited sample, SN\,2024iss demonstrates that comparing the SSA-derived shock velocity with the theoretical self-similar solution can serve as an effective method to test the possible existence of a confined CSM around a Type~IIb progenitor.

\section{Conclusion}\label{sec:5}
We conducted JVN single-baseline VLBI monitoring of the Type~IIb SN\,2024iss at 6.9 and 8.4\>GHz for approximately one year after its discovery. The SN was detected at 10 and 23 days after the explosion, while the subsequent observations yielded non-detections. Based on the estimated peak radio luminosity and peak time, SN\,2024iss exhibits radio properties highly comparable to those of the cIIb events. Using a SSA model, we estimate a progenitor mass-loss rate of $\dot{M} \approx 2.5 \times 10^{-6}\>M_{\odot}\>{\rm yr^{-1}}$ for a compact progenitor case (wind velocity of $100\>{\rm km\>s^{-1}}$).

Furthermore, our SSA analysis yields a mean expansion velocity of $V_{\rm sh} \approx 3.3 \times 10^4\>{\rm km\>s^{-1}}$, which exceeds the theoretical shock velocity derived from the self-similar solution by a factor of $\sim 2.4$. Even using the conservative upper-bound peak time, the velocity ratio remains $\gtrsim 1.7$. This velocity excess may be explained if the shock was accelerated as it emerged from a confined, dense CSM surrounding the progenitor, pointing to a complex, non-steady mass-loss history shortly before the explosion.

Although examples of confined CSM in Type~IIb SNe are still limited, SN\,2024iss demonstrates that comparing the SSA-derived velocity with the self-similar solution by radio observations serves as a diagnostic for such CSM structures. The formation of a confined CSM around a highly stripped Type~IIb progenitor, similar to those widely recognized in red supergiants for Type~IIP SNe, implies that the mechanism driving late-stage mass-loss enhancement might be triggered by the mere existence of a hydrogen envelope, largely independent of its mass.

\begin{ack}
This work was partially supported by the Inter-university collaborative project, the Japanese VLBI Network (JVN) of the National Astronomical Observatory of Japan. We are grateful to the JVN staff members who helped operate the telescopes and data transfer. YI was supported by the Japan Society for the Promotion of Science (JSPS) KAKENHI grant JP23K13151. KN was supported by the JSPS KAKENHI grant JP15H00784. KM acknowledges support from JSPS KAKENHI grants JP23H04894, JP24KK0070 and JP24H01810. TM was supported by the JSPS KAKENHI grant JP26KJ0055, the National Science and Technology Council, Taiwan, under grant Nos. MOST 110-2112-M-001-068-MY3, 113-2112-M-001-028-, and NSTC114-2112-M-001-012-, and the Academia Sinica, Taiwan, under a career development award under grant No. AS-CDA-111-M04.
\end{ack}








\bibliographystyle{aasjournal}
\bibliography{reference}

@ARTICLE{VanDyk13,
       author = {{Van Dyk}, Schuyler D. and {Zheng}, WeiKang and {Clubb}, Kelsey I. and {Filippenko}, Alexei V. and {Cenko}, S. Bradley and {Smith}, Nathan and {Fox}, Ori D. and {Kelly}, Patrick L. and {Shivvers}, Isaac and {Ganeshalingam}, Mohan},
        title = "{The Progenitor of Supernova 2011dh has Vanished}",
      journal = {\apjl},
         year = 2013,
        month = aug,
       volume = {772},
       number = {2},
          eid = {L32},
        pages = {L32},
          doi = {10.1088/2041-8205/772/2/L32},
archivePrefix = {arXiv},
       eprint = {1305.3436},
 primaryClass = {astro-ph.SR},
       adsurl = {https://ui.adsabs.harvard.edu/abs/2013ApJ...772L..32V}
}

@ARTICLE{Maund11,
       author = {{Maund}, J.~R. and {Fraser}, M. and {Ergon}, M. and {Pastorello}, A. and {Smartt}, S.~J. and {Sollerman}, J. and {Benetti}, S. and {Botticella}, M.-T. and {Bufano}, F. and {Danziger}, I.~J. and {Kotak}, R. and {Magill}, L. and {Stephens}, A.~W. and {Valenti}, S.},
        title = "{The Yellow Supergiant Progenitor of the Type II Supernova 2011dh in M51}",
      journal = {\apjl},
         year = 2011,
        month = oct,
       volume = {739},
       number = {2},
          eid = {L37},
        pages = {L37},
          doi = {10.1088/2041-8205/739/2/L37},
archivePrefix = {arXiv},
       eprint = {1106.2565},
 primaryClass = {astro-ph.SR},
       adsurl = {https://ui.adsabs.harvard.edu/abs/2011ApJ...739L..37M}
}

@ARTICLE{Soderberg12,
       author = {{Soderberg}, A.~M. and {Margutti}, R. and {Zauderer}, B.~A. and {Krauss}, M. and {Katz}, B. and {Chomiuk}, L. and {Dittmann}, J.~A. and {Nakar}, E. and {Sakamoto}, T. and {Kawai}, N. and {Hurley}, K. and {Barthelmy}, S. and {Toizumi}, T. and {Morii}, M. and {Chevalier}, R.~A. and {Gurwell}, M. and {Petitpas}, G. and {Rupen}, M. and {Alexander}, K.~D. and {Levesque}, E.~M. and {Fransson}, C. and {Brunthaler}, A. and {Bietenholz}, M.~F. and {Chugai}, N. and {Grindlay}, J. and {Copete}, A. and {Connaughton}, V. and {Briggs}, M. and {Meegan}, C. and {von Kienlin}, A. and {Zhang}, X. and {Rau}, A. and {Golenetskii}, S. and {Mazets}, E. and {Cline}, T.},
        title = "{Panchromatic Observations of SN 2011dh Point to a Compact Progenitor Star}",
      journal = {\apj},
         year = 2012,
        month = jun,
       volume = {752},
       number = {2},
          eid = {78},
        pages = {78},
          doi = {10.1088/0004-637X/752/2/78},
archivePrefix = {arXiv},
       eprint = {1107.1876},
 primaryClass = {astro-ph.HE},
       adsurl = {https://ui.adsabs.harvard.edu/abs/2012ApJ...752...78S}
}

@ARTICLE{Arcavi11,
       author = {{Arcavi}, Iair and {Gal-Yam}, Avishay and {Yaron}, Ofer and {Sternberg}, Assaf and {Rabinak}, Itay and {Waxman}, Eli and {Kasliwal}, Mansi M. and {Quimby}, Robert M. and {Ofek}, Eran O. and {Horesh}, Assaf and {Kulkarni}, Shrinivas R. and {Filippenko}, Alexei V. and {Silverman}, Jeffrey M. and {Cenko}, S. Bradley and {Li}, Weidong and {Bloom}, Joshua S. and {Sullivan}, Mark and {Nugent}, Peter E. and {Poznanski}, Dovi and {Gorbikov}, Evgeny and {Fulton}, Benjamin J. and {Howell}, D. Andrew and {Bersier}, David and {Riou}, Amedee and {Lamotte-Bailey}, Stephane and {Griga}, Thomas and {Cohen}, Judith G. and {Hachinger}, Stephan and {Polishook}, David and {Xu}, Dong and {Ben-Ami}, Sagi and {Manulis}, Ilan and {Walker}, Emma S. and {Maguire}, Kate and {Pan}, Yen-Chen and {Matheson}, Thomas and {Mazzali}, Paolo A. and {Pian}, Elena and {Fox}, Derek B. and {Gehrels}, Neil and {Law}, Nicholas and {James}, Philip and {Marchant}, Jonathan M. and {Smith}, Robert J. and {Mottram}, Chris J. and {Barnsley}, Robert M. and {Kandrashoff}, Michael T. and {Clubb}, Kelsey I.},
        title = "{SN 2011dh: Discovery of a Type IIb Supernova from a Compact Progenitor in the Nearby Galaxy M51}",
      journal = {\apjl},
         year = 2011,
        month = dec,
       volume = {742},
       number = {2},
          eid = {L18},
        pages = {L18},
          doi = {10.1088/2041-8205/742/2/L18},
archivePrefix = {arXiv},
       eprint = {1106.3551},
 primaryClass = {astro-ph.CO},
       adsurl = {https://ui.adsabs.harvard.edu/abs/2011ApJ...742L..18A}
}

@ARTICLE{Maeda14,
       author = {{Maeda}, Keiichi and {Katsuda}, Satoru and {Bamba}, Aya and {Terada}, Yukikatsu and {Fukazawa}, Yasushi},
        title = "{Long-lasting X-Ray Emission from Type IIb Supernova 2011dh and Mass-loss History of the Yellow Supergiant Progenitor}",
      journal = {\apj},
         year = 2014,
        month = apr,
       volume = {785},
       number = {2},
          eid = {95},
        pages = {95},
          doi = {10.1088/0004-637X/785/2/95},
archivePrefix = {arXiv},
       eprint = {1403.2455},
 primaryClass = {astro-ph.HE},
       adsurl = {https://ui.adsabs.harvard.edu/abs/2014ApJ...785...95M}
}

@ARTICLE{Sasaki12,
       author = {{Sasaki}, M. and {Ducci}, L.},
        title = "{Observations of the post-shock break-out emission of SN 2011dh with XMM-Newton}",
      journal = {\aap},
         year = 2012,
        month = oct,
       volume = {546},
          eid = {A80},
        pages = {A80},
          doi = {10.1051/0004-6361/201220200},
archivePrefix = {arXiv},
       eprint = {1209.4720},
 primaryClass = {astro-ph.HE},
       adsurl = {https://ui.adsabs.harvard.edu/abs/2012A&A...546A..80S}
}

@ARTICLE{Matsuoka25,
       author = {{Matsuoka}, Tomoki and {Maeda}, Keiichi and {Chen}, Ke-Jung},
        title = "{Inferring a Dense Confined Circumstellar Medium around Supernova Progenitors via Long-term Hydrodynamical Evolution}",
      journal = {\apjl},
         year = 2025,
        month = aug,
       volume = {988},
       number = {2},
          eid = {L53},
        pages = {L53},
          doi = {10.3847/2041-8213/adef08},
archivePrefix = {arXiv},
       eprint = {2504.14255},
 primaryClass = {astro-ph.HE},
       adsurl = {https://ui.adsabs.harvard.edu/abs/2025ApJ...988L..53M}
}

@ARTICLE{Folatelli15,
       author = {{Folatelli}, Gast{\'o}n and {Bersten}, Melina C. and {Kuncarayakti}, Hanindyo and {Benvenuto}, Omar G. and {Maeda}, Keiichi and {Nomoto}, Ken'ichi},
        title = "{The Progenitor of the Type IIb SN 2008ax Revisited}",
      journal = {\apj},
         year = 2015,
        month = oct,
       volume = {811},
       number = {2},
          eid = {147},
        pages = {147},
          doi = {10.1088/0004-637X/811/2/147},
archivePrefix = {arXiv},
       eprint = {1509.01588},
 primaryClass = {astro-ph.SR},
       adsurl = {https://ui.adsabs.harvard.edu/abs/2015ApJ...811..147F}
}

@ARTICLE{Smith08,
       author = {{Smith}, Nathan and {Conti}, Peter S.},
        title = "{On the Role of the WNH Phase in the Evolution of Very Massive Stars: Enabling the LBV Instability with Feedback}",
      journal = {\apj},
         year = 2008,
        month = jun,
       volume = {679},
       number = {2},
        pages = {1467-1477},
          doi = {10.1086/586885},
archivePrefix = {arXiv},
       eprint = {0802.1742},
 primaryClass = {astro-ph},
       adsurl = {https://ui.adsabs.harvard.edu/abs/2008ApJ...679.1467S}
}

@ARTICLE{Claeys11,
       author = {{Claeys}, J.~S.~W. and {de Mink}, S.~E. and {Pols}, O.~R. and {Eldridge}, J.~J. and {Baes}, M.},
        title = "{Binary progenitor models of type IIb supernovae}",
      journal = {\aap},
         year = 2011,
        month = apr,
       volume = {528},
          eid = {A131},
        pages = {A131},
          doi = {10.1051/0004-6361/201015410},
archivePrefix = {arXiv},
       eprint = {1102.1732},
 primaryClass = {astro-ph.SR},
       adsurl = {https://ui.adsabs.harvard.edu/abs/2011A&A...528A.131C}
}

@ARTICLE{Folatelli14,
       author = {{Folatelli}, Gast{\'o}n and {Bersten}, Melina C. and {Benvenuto}, Omar G. and {Van Dyk}, Schuyler D. and {Kuncarayakti}, Hanindyo and {Maeda}, Keiichi and {Nozawa}, Takaya and {Nomoto}, Ken'ichi and {Hamuy}, Mario and {Quimby}, Robert M.},
        title = "{A Blue Point Source at the Location of Supernova 2011dh}",
      journal = {\apjl},
         year = 2014,
        month = oct,
       volume = {793},
       number = {2},
          eid = {L22},
        pages = {L22},
          doi = {10.1088/2041-8205/793/2/L22},
archivePrefix = {arXiv},
       eprint = {1409.0700},
 primaryClass = {astro-ph.SR},
       adsurl = {https://ui.adsabs.harvard.edu/abs/2014ApJ...793L..22F}
}

@ARTICLE{Benvenuto13,
       author = {{Benvenuto}, Omar G. and {Bersten}, Melina C. and {Nomoto}, Ken'ichi},
        title = "{A Binary Progenitor for the Type IIb Supernova 2011dh in M51}",
      journal = {\apj},
         year = 2013,
        month = jan,
       volume = {762},
       number = {2},
          eid = {74},
        pages = {74},
          doi = {10.1088/0004-637X/762/2/74},
archivePrefix = {arXiv},
       eprint = {1207.5807},
 primaryClass = {astro-ph.SR},
       adsurl = {https://ui.adsabs.harvard.edu/abs/2013ApJ...762...74B}
}

@ARTICLE{Nomoto95,
       author = {{Nomoto}, K.~I. and {Iwamoto}, K. and {Suzuki}, T.},
        title = "{The evolution and explosion of massive binary stars and Type Ib-Ic-IIb-IIL supernovae.}",
      journal = {\physrep},
         year = 1995,
        month = may,
       volume = {256},
       number = {1},
        pages = {173-191},
          doi = {10.1016/0370-1573(94)00107-E},
       adsurl = {https://ui.adsabs.harvard.edu/abs/1995PhR...256..173N}
}

@ARTICLE{Bersten12,
       author = {{Bersten}, Melina C. and {Benvenuto}, Omar G. and {Nomoto}, Ken'ichi and {Ergon}, Mattias and {Folatelli}, Gast{\'o}n and {Sollerman}, Jesper and {Benetti}, Stefano and {Botticella}, Maria Teresa and {Fraser}, Morgan and {Kotak}, Rubina and {Maeda}, Keiichi and {Ochner}, Paolo and {Tomasella}, Lina},
        title = "{The Type IIb Supernova 2011dh from a Supergiant Progenitor}",
      journal = {\apj},
         year = 2012,
        month = sep,
       volume = {757},
       number = {1},
          eid = {31},
        pages = {31},
          doi = {10.1088/0004-637X/757/1/31},
archivePrefix = {arXiv},
       eprint = {1207.5975},
 primaryClass = {astro-ph.HE},
       adsurl = {https://ui.adsabs.harvard.edu/abs/2012ApJ...757...31B}
}

@ARTICLE{VanDyk14,
       author = {{Van Dyk}, Schuyler D. and {Zheng}, WeiKang and {Fox}, Ori D. and {Cenko}, S. Bradley and {Clubb}, Kelsey I. and {Filippenko}, Alexei V. and {Foley}, Ryan J. and {Miller}, Adam A. and {Smith}, Nathan and {Kelly}, Patrick L. and {Lee}, William H. and {Ben-Ami}, Sagi and {Gal-Yam}, Avishay},
        title = "{The Type IIb Supernova 2013df and its Cool Supergiant Progenitor}",
      journal = {\aj},
         year = 2014,
        month = feb,
       volume = {147},
       number = {2},
          eid = {37},
        pages = {37},
          doi = {10.1088/0004-6256/147/2/37},
archivePrefix = {arXiv},
       eprint = {1312.3984},
 primaryClass = {astro-ph.SR},
       adsurl = {https://ui.adsabs.harvard.edu/abs/2014AJ....147...37V}
}

@ARTICLE{VanDyk11,
       author = {{Van Dyk}, Schuyler D. and {Li}, Weidong and {Cenko}, S. Bradley and {Kasliwal}, Mansi M. and {Horesh}, Assaf and {Ofek}, Eran O. and {Kraus}, Adam L. and {Silverman}, Jeffrey M. and {Arcavi}, Iair and {Filippenko}, Alexei V. and {Gal-Yam}, Avishay and {Quimby}, Robert M. and {Kulkarni}, Shrinivas R. and {Yaron}, Ofer and {Polishook}, David},
        title = "{The Progenitor of Supernova 2011dh/PTF11eon in Messier 51}",
      journal = {\apjl},
         year = 2011,
        month = nov,
       volume = {741},
       number = {2},
          eid = {L28},
        pages = {L28},
          doi = {10.1088/2041-8205/741/2/L28},
archivePrefix = {arXiv},
       eprint = {1106.2897},
 primaryClass = {astro-ph.CO},
       adsurl = {https://ui.adsabs.harvard.edu/abs/2011ApJ...741L..28V}
}

@ARTICLE{Aldering94,
       author = {{Aldering}, G. and {Humphreys}, R.~M. and {Richmond}, M.},
        title = "{SN 1993J: The Optical Properties of its Progenitor}",
      journal = {\aj},
         year = 1994,
        month = feb,
       volume = {107},
        pages = {662},
          doi = {10.1086/116886},
       adsurl = {https://ui.adsabs.harvard.edu/abs/1994AJ....107..662A}
}

@ARTICLE{Nomoto93,
       author = {{Nomoto}, K. and {Suzuki}, T. and {Shigeyama}, T. and {Kumagai}, S. and {Yamaoka}, H. and {Saio}, H.},
        title = "{A type IIb model for supernova 1993J}",
      journal = {\nat},
         year = 1993,
        month = aug,
       volume = {364},
       number = {6437},
        pages = {507-509},
          doi = {10.1038/364507a0},
       adsurl = {https://ui.adsabs.harvard.edu/abs/1993Natur.364..507N}
}

@ARTICLE{Filippenko93,
       author = {{Filippenko}, Alexei V. and {Matheson}, Thomas and {Ho}, Luis C.},
        title = "{The ``Type IIb'' Supernova 1993J in M81: A Close Relative of Type Ib Supernovae}",
      journal = {\apjl},
         year = 1993,
        month = oct,
       volume = {415},
        pages = {L103},
          doi = {10.1086/187043},
       adsurl = {https://ui.adsabs.harvard.edu/abs/1993ApJ...415L.103F}
}

@ARTICLE{Yoon10,
       author = {{Yoon}, S.-C. and {Woosley}, S.~E. and {Langer}, N.},
        title = "{Type Ib/c Supernovae in Binary Systems. I. Evolution and Properties of the Progenitor Stars}",
      journal = {\apj},
         year = 2010,
        month = dec,
       volume = {725},
       number = {1},
        pages = {940-954},
          doi = {10.1088/0004-637X/725/1/940},
archivePrefix = {arXiv},
       eprint = {1004.0843},
 primaryClass = {astro-ph.SR},
       adsurl = {https://ui.adsabs.harvard.edu/abs/2010ApJ...725..940Y}
}

@ARTICLE{Podsiadlowski92,
       author = {{Podsiadlowski}, Ph. and {Joss}, P.~C. and {Hsu}, J.~J.~L.},
        title = "{Presupernova Evolution in Massive Interacting Binaries}",
      journal = {\apj},
         year = 1992,
        month = may,
       volume = {391},
        pages = {246},
          doi = {10.1086/171341},
       adsurl = {https://ui.adsabs.harvard.edu/abs/1992ApJ...391..246P}
}

@ARTICLE{Bufano14,
       author = {{Bufano}, F. and {Pignata}, G. and {Bersten}, M. and {Mazzali}, P.~A. and {Ryder}, S.~D. and {Margutti}, R. and {Milisavljevic}, D. and {Morelli}, L. and {Benetti}, S. and {Cappellaro}, E. and {Gonzalez-Gaitan}, S. and {Romero-Ca{\~n}izales}, C. and {Stritzinger}, M. and {Walker}, E.~S. and {Anderson}, J.~P. and {Contreras}, C. and {de Jaeger}, T. and {F{\"o}rster}, F. and {Gutierrez}, C. and {Hamuy}, M. and {Hsiao}, E. and {Morrell}, N. and {Olivares E.}, F. and {Paillas}, E. and {Parker}, S. and {Pian}, E. and {Pickering}, T.~E. and {Sanders}, N. and {Stockdale}, C. and {Turatto}, M. and {Valenti}, S. and {Fesen}, R.~A. and {Maza}, J. and {Nomoto}, K. and {Phillips}, M.~M. and {Soderberg}, A.},
        title = "{SN 2011hs: a fast and faint Type IIb supernova from a supergiant progenitor}",
      journal = {\mnras},
         year = 2014,
        month = apr,
       volume = {439},
       number = {2},
        pages = {1807-1828},
          doi = {10.1093/mnras/stu065},
archivePrefix = {arXiv},
       eprint = {1401.2368},
 primaryClass = {astro-ph.SR},
       adsurl = {https://ui.adsabs.harvard.edu/abs/2014MNRAS.439.1807B}
}

@ARTICLE{Maeda15,
       author = {{Maeda}, K. and {Hattori}, T. and {Milisavljevic}, D. and {Folatelli}, G. and {Drout}, M.~R. and {Kuncarayakti}, H. and {Margutti}, R. and {Kamble}, A. and {Soderberg}, A. and {Tanaka}, M. and {Kawabata}, M. and {Kawabata}, K.~S. and {Yamanaka}, M. and {Nomoto}, K. and {Kim}, J.~H. and {Simon}, J.~D. and {Phillips}, M.~M. and {Parrent}, J. and {Nakaoka}, T. and {Moriya}, T.~J. and {Suzuki}, A. and {Takaki}, K. and {Ishigaki}, M. and {Sakon}, I. and {Tajitsu}, A. and {Iye}, M.},
        title = "{Type IIb Supernova 2013df Entering into an Interaction Phase: A Link between the Progenitor and the Mass Loss}",
      journal = {\apj},
         year = 2015,
        month = jul,
       volume = {807},
       number = {1},
          eid = {35},
        pages = {35},
          doi = {10.1088/0004-637X/807/1/35},
archivePrefix = {arXiv},
       eprint = {1504.06668},
 primaryClass = {astro-ph.SR},
       adsurl = {https://ui.adsabs.harvard.edu/abs/2015ApJ...807...35M}
}

@ARTICLE{Chen26,
       author = {{Chen}, Liyang and {Wang}, Xiaofeng and {Wu}, Qinyu and {Andrews}, Moira and {Farah}, Joseph and {Ochner}, Paolo and {Reguitti}, Andrea and {Brink}, Thomas G. and {Zhang}, Jujia and {Song}, Cuiying and {Liu}, Jialian and {Filippenko}, Alexei V. and {Sand}, David J. and {Albanese}, Irene and {Alexander}, Kate D. and {Andrews}, Jennifer and {Bostroem}, K. Azalee and {Cai}, Yongzhi and {Christy}, Collin and {Esamdin}, Ali and {Farina}, Andrea and {Franz}, Noah and {Howell}, D. Andrew and {Hsu}, Brian and {Hu}, Maokai and {Iskandar}, Abdusamatjan and {Li}, Liping and {Li}, Gaici and {Li}, Dongyue and {Li}, Wenxiong and {Liu}, Jinzhong and {McCully}, Curtis and {Newsome}, Megan and {Ni}, Yuan Qi and {Pastorello}, Andrea and {Padilla Gonzalez}, Estefania and {Pearson}, Jeniveve and {Peng}, Haowei and {Ransome}, Conor and {Shrestha}, Manisha and {Smith}, Nathan and {Subrayan}, Bhagya and {Terreran}, Giacomo and {Valerin}, Giorgio and {Vink{\'o}}, J. and {Vasylyev}, Sergiy S. and {Wang}, Letian and {Wang}, Zhenyu and {Wang}, Hao and {Wheeler}, J. Craig and {Wynn}, Kathryn and {Xiang}, Danfeng and {Yan}, Shengyu and {Yuan}, Weimin and {Zhang}, Juan and {Zheng}, WeiKang and {Zhang}, Yu},
        title = "{SN 2024iss: A Double-peaked Type IIb Supernova with Evidence of Circumstellar Interaction}",
      journal = {arXiv e-prints},
         year = 2025,
        month = oct,
          eid = {arXiv:2510.22997},
        pages = {arXiv:2510.22997},
          doi = {10.48550/arXiv.2510.22997},
archivePrefix = {arXiv},
       eprint = {2510.22997},
 primaryClass = {astro-ph.HE},
       adsurl = {https://ui.adsabs.harvard.edu/abs/2025arXiv251022997C}
}

@ARTICLE{Chevalier10,
       author = {{Chevalier}, Roger A. and {Soderberg}, Alicia M.},
        title = "{Type IIb Supernovae with Compact and Extended Progenitors}",
      journal = {\apjl},
         year = 2010,
        month = mar,
       volume = {711},
       number = {1},
        pages = {L40-L43},
          doi = {10.1088/2041-8205/711/1/L40},
archivePrefix = {arXiv},
       eprint = {0911.3408},
 primaryClass = {astro-ph.HE},
       adsurl = {https://ui.adsabs.harvard.edu/abs/2010ApJ...711L..40C}
}

@ARTICLE{Fransson96,
       author = {{Fransson}, Claes and {Lundqvist}, Peter and {Chevalier}, Roger A.},
        title = "{Circumstellar Interaction in SN 1993J}",
      journal = {\apj},
         year = 1996,
        month = apr,
       volume = {461},
        pages = {993},
          doi = {10.1086/177119},
       adsurl = {https://ui.adsabs.harvard.edu/abs/1996ApJ...461..993F}
}

@ARTICLE{Chevalier06,
       author = {{Chevalier}, Roger A. and {Fransson}, Claes},
        title = "{Circumstellar Emission from Type Ib and Ic Supernovae}",
      journal = {\apj},
         year = 2006,
        month = nov,
       volume = {651},
       number = {1},
        pages = {381-391},
          doi = {10.1086/507606},
archivePrefix = {arXiv},
       eprint = {astro-ph/0607196},
 primaryClass = {astro-ph},
       adsurl = {https://ui.adsabs.harvard.edu/abs/2006ApJ...651..381C}
}

@ARTICLE{Srivastav24,
       author = {{Srivastav}, S. and {Fulton}, M. and {Nicholl}, M. and {Angus}, C.~R. and {Smith}, K.~W. and {Young}, D.~R. and {Moore}, T. and {Sim}, S.~A. and {Smartt}, S.~J. and {Chen}, T.~W.},
        title = "{OxQUB Transient Classification Report for 2024-06-21}",
      journal = {Transient Name Server Classification Report},
         year = 2024,
        month = jun,
       volume = {2024-2046},
        pages = {1},
       adsurl = {https://ui.adsabs.harvard.edu/abs/2024TNSCR2046....1S}
}

@ARTICLE{Iwata25,
       author = {{Iwata}, Yuhei and {Akimoto}, Masanori and {Matsuoka}, Tomoki and {Maeda}, Keiichi and {Yonekura}, Yoshinori and {Tominaga}, Nozomu and {Moriya}, Takashi J. and {Fujisawa}, Kenta and {Niinuma}, Kotaro and {Yoon}, Sung-Chul and {Lee}, Jae-Joon and {Jung}, Taehyun and {Byun}, Do-Young},
        title = "{Radio Follow-up Observations of SN 2023ixf by Japanese and Korean Very Long Baseline Interferometers}",
      journal = {\apj},
         year = 2025,
        month = jan,
       volume = {978},
       number = {2},
          eid = {138},
        pages = {138},
          doi = {10.3847/1538-4357/ad9a62},
archivePrefix = {arXiv},
       eprint = {2411.07542},
 primaryClass = {astro-ph.HE},
       adsurl = {https://ui.adsabs.harvard.edu/abs/2025ApJ...978..138I}
}

@article{Weiler02,
	adsurl = {https://ui.adsabs.harvard.edu/abs/2002ARA&A..40..387W},
	author = {{Weiler}, Kurt W. and {Panagia}, Nino and {Montes}, M. J. and {Sramek}, Richard A.},
	doi = {10.1146/annurev.astro.40.060401.093947},
	journal = {Annual Review of Astronomy and Astrophysics},
	month = sep,
	pages = {387-438},
	title = {{Radio Emission from Supernovae and Gamma-Ray Bursters}},
	volume = {40},
	year = 2002}

@article{Perley17,
	adsurl = {https://ui.adsabs.harvard.edu/abs/2017ApJS..230....7P},
	author = {{Perley}, Daniel A. and {Butler}, Bryan J.},
	doi = {10.3847/1538-4365/aa6df9},
	journal = {\apjs},
	month = may,
	number = {1},
	pages = {7},
	title = {{An Accurate Flux Density Scale from 1 to 50 GHz}},
	volume = {230},
	year = 2017}

@article{ONeill24,
	adsurl = {https://ui.adsabs.harvard.edu/abs/2024TNSAN.128....1O},
	author = {{O'Neill}, D. and {Godson}, B. and {Killestein}, T. and {Kotak}, R. and {Kuncarayakti}, H. and {Pursiainen}, M. and {Ackley}, K. and {Dyer}, M. and {Jim{\'e}nez-Ibarra}, F. and {Lyman}, J. and {Ulaczyk}, K. and {Steeghs}, D. and {Galloway}, D. and {Dhillon}, V. and {O'Brien}, P. and {Ramsay}, G. and {Noysena}, K. and {Breton}, R. and {Nuttall}, L. and {Pall{\'e}}, E. and {Pollacco}, D. and {Kumar}, A.},
	journal = {Transient Name Server AstroNote},
	month = may,
	pages = {1},
	title = {{GOTO discovery of a nearby bright transient at 18.7 Mpc.}},
	volume = {128},
	year = 2024}

@article{Sfaradi24,
	adsurl = {https://ui.adsabs.harvard.edu/abs/2024ATel16634....1S},
	author = {{Sfaradi}, Itai and {Horesh}, Assaf and {Bright}, Joe and {Farah}, Wael and {Pollak}, Alexander and {Siemion}, Andrew},
	journal = {The Astronomer's Telegram},
	month = may,
	pages = {1},
	title = {{AMI-LA 15.5 GHz Radio Detection of SN~2024iss}},
	volume = {16634},
	year = 2024}

@article{Bright24,
	adsurl = {https://ui.adsabs.harvard.edu/abs/2024ATel16644....1B},
	author = {{Bright}, Joe and {Farah}, Wael and {Pollak}, Alexander and {Siemion}, Andrew and {Sfaradi}, Itai and {Horesh}, Assaf},
	journal = {The Astronomer's Telegram},
	month = jun,
	pages = {1},
	title = {{Allen Telescope Array 7 GHz Radio Detection of SN2024iss}},
	volume = {16644},
	year = 2024}

@article{Yamanaka25,
	adsurl = {https://ui.adsabs.harvard.edu/abs/2025PASJ...77L..31Y},
	archiveprefix = {arXiv},
	author = {{Yamanaka}, Masayuki and {Nagayama}, Takahiro and {Horikiri}, Tsukiha},
	doi = {10.1093/pasj/psaf020},
	eprint = {2503.05054},
	journal = {\pasj},
	month = jun,
	number = {3},
	pages = {L31-L35},
	primaryclass = {astro-ph.HE},
	title = {{SN 2024iss: Double-peaked light curves and implications for a yellow supergiant progenitor}},
	volume = {77},
	year = 2025}

@article{Matsuoka19,
	adsurl = {https://ui.adsabs.harvard.edu/abs/2019ApJ...885...41M},
	archiveprefix = {arXiv},
	author = {{Matsuoka}, Tomoki and {Maeda}, Keiichi and {Lee}, Shiu-Hang and {Yasuda}, Haruo},
	doi = {10.3847/1538-4357/ab4421},
	eid = {41},
	eprint = {1909.05874},
	journal = {\apj},
	month = nov,
	number = {1},
	pages = {41},
	primaryclass = {astro-ph.HE},
	title = {{Radio Emission from Supernovae in the Very Early Phase: Implications for the Dynamical Mass Loss of Massive Stars}},
	volume = {885},
	year = 2019}

@article{Chevalier82a,
	adsurl = {https://ui.adsabs.harvard.edu/abs/1982ApJ...258..790C},
	author = {{Chevalier}, R.~A.},
	doi = {10.1086/160126},
	journal = {\apj},
	month = jul,
	pages = {790-797},
	title = {{Self-similar solutions for the interaction of stellar ejecta with an external medium.}},
	volume = {258},
	year = 1982}

@ARTICLE{Chevalier82b,
       author = {{Chevalier}, R.~A.},
        title = "{The radio and X-ray emission from type II supernovae.}",
      journal = {\apj},
         year = 1982,
        month = aug,
       volume = {259},
        pages = {302-310},
          doi = {10.1086/160167},
       adsurl = {https://ui.adsabs.harvard.edu/abs/1982ApJ...259..302C}
}

@ARTICLE{Nayana22,
       author = {{Nayana}, A.~J. and {Chandra}, Poonam and {Krishna}, Anoop and {Anupama}, G.~C.},
        title = "{Radio Evolution of a Type IIb Supernova SN 2016gkg}",
      journal = {\apj},
         year = 2022,
        month = aug,
       volume = {934},
       number = {2},
          eid = {186},
        pages = {186},
          doi = {10.3847/1538-4357/ac7c1e},
archivePrefix = {arXiv},
       eprint = {2206.12103},
 primaryClass = {astro-ph.HE},
       adsurl = {https://ui.adsabs.harvard.edu/abs/2022ApJ...934..186N}
}

@article{Maeda23,
	adsurl = {https://ui.adsabs.harvard.edu/abs/2023ApJ...945L...3M},
	archiveprefix = {arXiv},
	author = {{Maeda}, Keiichi and {Michiyama}, Tomonari and {Chandra}, Poonam and {Ryder}, Stuart and {Kuncarayakti}, Hanindyo and {Hiramatsu}, Daichi and {Imanishi}, Masatoshi},
	doi = {10.3847/2041-8213/acb25e},
	eid = {L3},
	eprint = {2301.07357},
	journal = {\apjl},
	month = mar,
	number = {1},
	pages = {L3},
	primaryclass = {astro-ph.HE},
	title = {{Resurrection of Type IIL Supernova 2018ivc: Implications for a Binary Evolution Sequence Connecting Hydrogen-rich and Hydrogen-poor Progenitors}},
	volume = {945},
	year = 2023}

@ARTICLE{Roming09,
       author = {{Roming}, P.~W.~A. and {Pritchard}, T.~A. and {Brown}, P.~J. and {Holland}, S.~T. and {Immler}, S. and {Stockdale}, C.~J. and {Weiler}, K.~W. and {Panagia}, N. and {Van Dyk}, S.~D. and {Hoversten}, E.~A. and {Milne}, P.~A. and {Oates}, S.~R. and {Russell}, B. and {Vandrevala}, C.},
        title = "{Multi-Wavelength Properties of the Type IIb SN 2008ax}",
      journal = {\apjl},
         year = 2009,
        month = oct,
       volume = {704},
       number = {2},
        pages = {L118-L123},
          doi = {10.1088/0004-637X/704/2/L118},
archivePrefix = {arXiv},
       eprint = {0909.0967},
 primaryClass = {astro-ph.HE},
       adsurl = {https://ui.adsabs.harvard.edu/abs/2009ApJ...704L.118R}
}

@article{Ryder04,
	adsurl = {https://ui.adsabs.harvard.edu/abs/2004MNRAS.349.1093R},
	archiveprefix = {arXiv},
	author = {{Ryder}, Stuart D. and {Sadler}, Elaine M. and {Subrahmanyan}, Ravi and {Weiler}, Kurt W. and {Panagia}, Nino and {Stockdale}, Christopher},
	doi = {10.1111/j.1365-2966.2004.07589.x},
	eprint = {astro-ph/0401135},
	journal = {\mnras},
	month = apr,
	number = {3},
	pages = {1093-1100},
	primaryclass = {astro-ph},
	title = {{Modulations in the radio light curve of the Type IIb supernova 2001ig: evidence for a Wolf-Rayet binary progenitor?}},
	volume = {349},
	year = 2004}

@ARTICLE{Weiler07,
       author = {{Weiler}, Kurt W. and {Williams}, Christopher L. and {Panagia}, Nino and {Stockdale}, Christopher J. and {Kelley}, Matthew T. and {Sramek}, Richard A. and {Van Dyk}, Schuyler D. and {Marcaide}, J.~M.},
        title = "{Long-Term Radio Monitoring of SN 1993J}",
      journal = {\apj},
         year = 2007,
        month = dec,
       volume = {671},
       number = {2},
        pages = {1959-1980},
          doi = {10.1086/523258},
archivePrefix = {arXiv},
       eprint = {0709.1136},
 primaryClass = {astro-ph},
       adsurl = {https://ui.adsabs.harvard.edu/abs/2007ApJ...671.1959W}
}

@article{Soderberg06,
	adsurl = {https://ui.adsabs.harvard.edu/abs/2006ApJ...651.1005S},
	archiveprefix = {arXiv},
	author = {{Soderberg}, A.~M. and {Chevalier}, R.~A. and {Kulkarni}, S.~R. and {Frail}, D.~A.},
	doi = {10.1086/507571},
	eprint = {astro-ph/0512413},
	journal = {\apj},
	month = nov,
	number = {2},
	pages = {1005-1018},
	primaryclass = {astro-ph},
	title = {{The Radio and X-Ray Luminous SN 2003bg and the Circumstellar Density Variations around Radio Supernovae}},
	volume = {651},
	year = 2006}

@article{Chevalier98,
	adsurl = {https://ui.adsabs.harvard.edu/abs/1998ApJ...499..810C},
	author = {{Chevalier}, Roger A.},
	doi = {10.1086/305676},
	journal = {\apj},
	month = may,
	number = {2},
	pages = {810-819},
	title = {{Synchrotron Self-Absorption in Radio Supernovae}},
	volume = {499},
	year = 1998}

@article{Bietenholz21,
	adsurl = {https://ui.adsabs.harvard.edu/abs/2021ApJ...908...75B},
	archiveprefix = {arXiv},
	author = {{Bietenholz}, M.~F. and {Bartel}, N. and {Argo}, M. and {Dua}, R. and {Ryder}, S. and {Soderberg}, A.},
	doi = {10.3847/1538-4357/abccd9},
	eid = {75},
	eprint = {2011.11737},
	journal = {\apj},
	month = feb,
	number = {1},
	pages = {75},
	primaryclass = {astro-ph.HE},
	title = {{The Radio Luminosity-risetime Function of Core-collapse Supernovae}},
	volume = {908},
	year = 2021}

@article{Fujisawa22,
	adsurl = {https://ui.adsabs.harvard.edu/abs/2022PASJ...74.1415F},
	author = {{Fujisawa}, Kenta and {Aoki}, Takahiro and {Kanazawa}, Sho and {Akimoto}, Masanori and {Ogura}, Tatsuya and {Mori}, Kurena and {Niinuma}, Kotaro and {Motogi}, Kazuhito and {Sawada-Satoh}, Satoko and {Takefuji}, Kazuhiro and {Ogawa}, Hideo and {Kimura}, Kimihiro and {Yonekura}, Yoshinori and {Honma}, Mareki},
	doi = {10.1093/pasj/psac078},
	journal = {\pasj},
	month = dec,
	number = {6},
	pages = {1415-1420},
	title = {{The Yamaguchi Interferometer}},
	volume = {74},
	year = 2022}

@article{Yonekura16,
	adsurl = {https://ui.adsabs.harvard.edu/abs/2016PASJ...68...74Y},
	author = {{Yonekura}, Yoshinori and {Saito}, Yu and {Sugiyama}, Koichiro and {Soon}, Kang Lou and {Momose}, Munetake and {Yokosawa}, Masayoshi and {Ogawa}, Hideo and {Kimura}, Kimihiro and {Abe}, Yasuhiro and {Nishimura}, Atsushi and {Hasegawa}, Yutaka and {Fujisawa}, Kenta and {Ohyama}, Tomoaki and {Kono}, Yusuke and {Miyamoto}, Yusuke and {Sawada-Satoh}, Satoko and {Kobayashi}, Hideyuki and {Kawaguchi}, Noriyuki and {Honma}, Mareki and {Shibata}, Katsunori M. and {Sato}, Katsuhisa and {Ueno}, Yuji and {Jike}, Takaaki and {Tamura}, Yoshiaki and {Hirota}, Tomoya and {Miyazaki}, Atsushi and {Niinuma}, Kotaro and {Sorai}, Kazuo and {Takaba}, Hiroshi and {Hachisuka}, Kazuya and {Kondo}, Tetsuro and {Sekido}, Mamoru and {Murata}, Yasuhiro and {Nakai}, Naomasa and {Omodaka}, Toshihiro},
	doi = {10.1093/pasj/psw045},
	eid = {74},
	journal = {\pasj},
	month = oct,
	number = {5},
	pages = {74},
	title = {{The Hitachi and Takahagi 32 m radio telescopes: Upgrade of the antennas from satellite communication to radio astronomy}},
	volume = {68},
	year = 2016}

\end{document}